\documentclass[aps,prl,preprint,groupedaddress]{revtex4-1}
\usepackage{amssymb}
\usepackage{latexsym}
\usepackage{amsmath}
\usepackage{amsthm}
\usepackage{amscd}
\usepackage{graphicx}
\newcommand{\bn}{\begin{enumerate}}
\newcommand{\en}{\end{enumerate}}

\newcommand{\p}{\partial}
\newcommand{\be}{\begin{equation}}
\newcommand{\ee}{\end{equation} \noindent}
\newcommand{\bs}{\begin{subequations}}
\newcommand{\es}{\end{subequations} \noindent}
\newcommand{\bea}{\begin{eqnarray}}
\newcommand{\eea}{\end{eqnarray} \noindent}
\newcommand{\ba}{\begin{array}}

\newcommand{\ea}{\end{array}}
\newcommand{\bi}{\begin{itemize}}
\newcommand{\ei}{\end{itemize}}
\newcommand{\nn}{\nonumber}

\newcommand{\om}{\omega}

\newcommand{\ph}{\varphi}

\newcommand{\ep}{\varepsilon}

\renewcommand{\Im}{\mathrm{Im}\,}

\begin{document}


\title{On extending the Painlev\'e test to the one-dimensional Vlasov
equation\\
\small{1. Case of one simple pole}}

\author{Piotr P. Goldstein}
\affiliation{Theoretical Physics Division, National Centre for
Nuclear Research, Ho\.za 69, 00-681 Warsaw Poland\\
$email\mathrm{:~piotr.goldstein@ncbj.gov.pl}$}

\begin{abstract}
An analysis of possible extension of the Painlev\'e test, to
encompass the one-dimensional Vlasov equation, is performed. The
extending requires a nontrivial generalization of the test. The
proposed singularity analysis provides classification of the
solutions possessing the Painlev\'e property by the order and
number of pole surfaces. The compatibility conditions for the
Laurent series have the form of an overdetermined system of 1st
order differential equations, which themselves need a
compatibility condition. This eventually leads to constraints
which implicitly yield a family of solutions. The complete
calculation is provided for the case of one simple order pole. The
solutions describe evolution of plasmas in a uniform electric
field.
\end{abstract}


\maketitle

\section{Introduction}
The Vlasov equation is the most important equation in the kinetic
description of plasmas. With smooth initial conditions, it
describes evolution of the one-particle (electron, ion)
distribution functions in high-temperature or low density plasmas.
Even its simplest one-dimensional electrostatic form is
nontrivial.

Only a few situations are known in which the nonlinear
Vlasov-Poisson system may be solved explicitly
\cite{BGK,Waterb,DrSe}. We try a new method which yields
classification of meromorphic solutions of the one-dimensional
Vlasov-Poisson system and leads to explicit solutions in some
simplest cases. The method is based on the singularity analysis.
It is similar to the well-known Painlev\'e (P-) test, which is
usually performed to distinguish between the integrable and
non-integrable ordinary and partial differential equations (ODE
and PDE), without actually solving the equations. The author was
motivated by the question whether the P-test may be generalized to
encompass the Vlasov equation and what useful outcome can be
provided by the possible generalization.

The one-dimensional (1D) one-component (electron) Vlasov-Poisson
system may be written as
\bea\label{1DV}
&&\p_t f + v\,\p_x f+a\,\p_{v}
f=0,\nn\\
&&\p_x a=\om_p^2\,\left(\int dv\,f-1\right),
\eea
where $f(x,v,t)$ is the distribution function of the electrons,
normalized to the length of the container $L$ in the one-particle
phase (postion-veolcity) space $(x,v)$, the subscripts $x,v,t$
denote differentiation; $a(x,t)=-e\, E(x,t)/m$ is the acceleration
which the electrons (of mass $m$ and charge $-e$) gain in the
self-consistent electric field $E$, while $\om_p$ is the electron
plasma frequency.
\be\label{PF}
\om_p^2= 4\pi n_0 e^2/m ~\text{(cgs),}\quad \text{or }  ~ \om_p^2
=n_0 e^2/(\ep_0 m) ~\text{(SI)}.
\ee
$n_0$, being the average electron density, $\ep_0$ the
permittivity of the vacuum. For simplicity, we have concentrated
on singly charged ions. Introduction of the ion charge does not
bring much new.

For three dimensional (3D) counterpart of \eqref{1DV}, the $x$ and
$v$ derivatives are replaced by 3D gradients in the respective
spaces, and their products with $v$ and $a$ by the respective dot
products.

The one-dimensional Vlasov equation may describe situations in
which the deviation from equilibrium is limited to one dimension.
It also describes the dynamics of the guiding centers of electron
gyroscopic motion in very strong constant magnetic fields.

The usual goal of the P-test: distinction between integrable and
non-integrable equations or between regular and chaotic behavior
of their solutions does not apply to the Vlasov equation, because
the equation definitely has chaotic solutions, even in one
dimension. In principle, the Vlasov equation can describe complete
dynamics of any classical system if the initial condition is a sum
of Dirac's deltas (then the equation is known as the Klimontovich
equation \cite{Klim}). Nevertheless the singularity analysis of
the equation may provide interesting information about the classes
of solutions which pass the integrability test.

A brief reminder: The Painlev\'e (P-) test is a shorthand name of
a test for the P-property. The property applies to ODE and PDE; it
is defined as absence of branch points and some essential singular
points (or manifolds for PDE) which are ``movable'', i.e. their
position depends on the initial or boundary conditions. The
allowed movable singularities are poles and such essential
singular points which do not introduce multivaluedness (see
\cite{Conte, CM} for details).

The classical P-test \cite{Sonia,ARS,WTC} relies on looking for
solutions, extended to complex independent variables,  in a form
of the Laurent series about a hypothetic movable pole (a movable
pole-manifold for PDE's). Then coefficients of the series are
calculated in a recurrent way.

The recurrence relations are linear algebraic equations or systems
of the linear algebraic equations. At the indices where these
equations are underdetermined (the determinant is zero and the
equations are consistent), the arbitrary constants (functions for
PDE's) provide the first integrals.

The classical P-test has its shortcomings (a problem already
raised by Painlev\'e \cite{Conte}). Firstly, it is local: it
examines the properties of the solution in the neighborhood of a
movable singularity. Therefore it provides only the necessary
condition for the P-property. A proof of convergence of the
Laurent series is difficult and therefore hardly ever done.
Secondly, the test does not examine solutions which have no poles:
a negative initial exponent is assumed in the Laurent series (this
exclusion of nonnegative initial exponents can be reduced to
nonnegative integers less than the degree of the equations). These
shortcomings will also be possessed by its modification for the
Vlasov equation. As the equation \eqref{1DV} is
integro-differential the test additionally requires an assumption
of global meromorphy of the solution in the velocity space. In
this aspect the Vlasov equation is more demanding than equations
containing integration over a parameter, analyzed in \cite{PG1}.

For the purpose of the analysis, we make several technical
simplifications. First, we analyze the system in the thermodynamic
limit, assuming that the plasma is neutral and cold at infinity.
Second, the description is non-relativistic, so that $v\in
\mathbb{R}$. Third, the analysis will be performed for the Vlasov
equation written in terms of distribution which is cumulative in
the position space. We define $F(x,v,t)$ by
\be
f(x,v,t)=F_x(x,v,t)+h(v),
\ee
where $h(v)$ represents the velocity distribution of the uniform
ion background. It is an analytic approximation of the Dirac
delta: the ions are assumed to constitute an immobile uniform
neutralizing background. This way $F$ may be interpreted as the
position-cumulative phase-space distribution of the net charge. In terms of
this distribution, we can write the Vlasov-Poisson system
\eqref{1DV} as a single equation
\be\label{totest}
F_{xt}+v\,F_{xx}+ \om_p^2\left(\int\limits_{-\infty}^\infty
F(x,v',t)dv'\right)\, [F_{xv}+h'(v)]=0.
\ee
This equation, extended to complex $x,v,t$ will be the object of
our singularity analysis.

Fourth, the equation of the movable singularity $\Phi(x,v,t)=0$ is
written as solved with respect to $v$, so that the expansion
variable in the Laurent series is
\be \Phi(x,v,t)=v - \phi(x,t)
\ee
(``Kruskal gauge'').

In the further calculation we assume the following notation:\\
The numerical subscripts number the consecutive terms of the
expansion. The numerical superscript adjacent to a symbol refers
to numbering of the poles of the solution in one, say upper,
complex half-plane of $v$. As before, the alphabetic subscripts
denote differentiation; they are separated by a comma from a
numerical subscript if they occur together.

While considering the behavior of the solution in the neighborhood
of a particular $j$-th pole surface $v-\phi^j(x,t)=0$, most of the
analysis will be performed in the Lagrangian variables $\xi^j,
\tau^j$ such that
\bea\label{Lagr}
&&x(\xi^j,\tau^j)=\xi^j+\int_0^{\tau^j} d\tau'\,\ph(\xi^j,\tau')\nn\\
&&t=\tau^j,
\eea
where $\ph^j(\xi^j,\tau^j)=\phi^j(x(\xi^j,\tau^j),\tau^j)$.

Description in terms of the Lagrangian variables has proven useful
in several works \cite{Infeld1,Infeld2}. It also significantly
simplifies our equations. In the Lagrangian variables the partial
derivatives are respectively
\bea
&&\p_{\xi^j} =x_{\xi^j}\p_x\nn\\
&&\p_{\tau^j}=\p_t+\ph^j\p_x
\eea
(although all the $\tau^j$ are equal to $t$, the derivatives with
respect to them differ, which justifies supplying the variable $\tau$  with the
superscript).

In the above equations, the integrations play the role of
antiderivatives in a neighborhood of a given pole, thus the
calculation remains local in time. For the purpose of deriving
special solutions, we have specified the lower limit of
integration as zero, which identifies $\xi$ with the initial value
of $x$. The replacement of zero by another value is trivial.


\section{The singularity analysis of the one-dimensional Vlasov equation}
\subsection{General properties}
In this section we extend the Painlev\'e test to the Vlasov
equation. In addition to the above mentioned assumption of global
meromorphy in $v$, the extension differs from its original in
several aspects. At each pole surface:
\bn\item In the lowest-order we obtain an
equation which we call the dispersion relation, instead of the
initial power and its coefficient. The ``Kruskal gauge''
$\Phi(x,v,t)=v-\phi(x,t)$ proves to be the most natural approach,
which immediately yields the integral in \eqref{1DV} by residua.
\item
Once the dispersion relation is satisfied, \textbf{all indices are
resonant}, so the sum of coefficients at the dominant power is
identically zero. Therefore the coefficients cannot be calculated
by solving algebraic equations like in the usual Painlev\'e test.
We get first-order differential equations instead.
\item
Still each of the coefficients $F_n$ of the
expansion can be calculated, by solving a relation
of order $n+1$.
\item
Although the recurrence relations are differential, one resonant
index occurs at each pole like in the classical P-test. For that
index, the coefficient at the calculated expansion coefficient
vanishes.
\item
The recurrence relations together with the dispersion relation
always constitute an overdetermined system. The compatibility
conditions impose constraints on the singularity manifold.
Determining them is one of the more difficult aspects of the task,
because the system, though linear in the expansion coefficients,
is nonlinear with respect to the manifold variable.
\en
\textbf{Details }

1. The Laurent series about the $j$-th pole of order $p_j$ reads
\be\label{expansion}
F(x,v,t)=[v-\phi^j(x,t)]^{p^j}\sum\limits_{n=0}^\infty
{F^j}_n(x,t)[v-\phi^j(x,t)]^n
\ee
This equation is substituted to \eqref{totest}. If we assume that
$p^j <0$ for each $j$, then we may perform the integration in
\eqref{totest} by residua. Let the contour be a half-circle based
on a large segment of the real axis and closed far away at
$\mathop{\mathfrak{\Im}} v
>0$ so as to encircle all the poles. We obtain
\be\label{disp} 2\pi
i\,\om_p^2\,{F^j}_{-p^j-1} =
\sum_{k=1}^m({\phi^k},_t+{\phi^k}{\phi^k},_x)=a(x,t)
\ee
The rationale behind calling it a dispersion relation lies in the
analogy with the Fourier analysis of linear PDE. Like in the
Fourier analysis (mutatis mutandis), the same expansion
\eqref{expansion} applied to a linear equation yields a multiplier
on each differentiation (here it is proportional to  $\Phi^{-1}$).
If we substituted such an expansion to a linear PDE, this would
result in a constraint on $\Phi$ (and its derivatives), analogous
to the relation between the frequency and wave vector(s) in the
Fourier analysis. However, the Vlasov equation is nonlinear, which
manifests itself in the dependence of the relation on the unknown
function through its expansion coefficient ${F^j}_{-p^j-1}$.

Equation \eqref{disp} (as well as the further equations) may be
written in a compact form if we use the Lagrangian variables
\eqref{Lagr}. The dispersion relation \eqref{disp} reads simply
\be\label{dispL} 2\pi
i\,\om_p^2\,{F^j}_{-p^j-1} =
\sum_{k=1}^m x^k_{\tau^k\tau^k}=a(x,t),
\ee
which shows why the coefficients ${F^j}_{-p^j-1}$ at a given
$(x,t)$ have the same value for all poles $j$. Simply the
coefficients are equal (up to a common constant) to the electron
acceleration caused by the electric field at the same point $x$.
The singularity manifolds undergo the same acceleration. This
makes them characteristic manifolds of the equation \eqref{1DV}
and would exclude their use in the classical P-test. Still they
remain useful in our scheme, especially for deriving new
solutions.

2. and 3. The recurrence formulae look identically for all the
poles, so we skip the pole numbering below. Unlike the traditional
P-test, here all terms in the algebraic equation for $F_n$ vanish
in the $n$-th relation, provided that the dispersion relation
\eqref{disp} is satisfied. Therefore, in order to obtain the
consecutive coefficients $F_k,~k=0,1,...$, we have to proceed to
the higher order. Instead of $F_n$, the $n\!-\!1$-th term of the
expansion, $F_{n-1}$ is given by an equation at the order $n$.
Moreover it is a first-order differential rather than algebraic
equation. It reads
\begin{align}\label{rec}
&-(p+n-1)\p_\tau\left(x_\xi^{-(p+n-1)}x_{\xi\tau}F_{n-1}\right)
=-\p_\tau\left(x_\xi^{-(p+n-1)}F_{n-2,\xi}\right)\nn\\&+\left(p\!+\!n\!-\!2\right)
x_\xi^{-(p+n-1)}\p_\xi\left(x_\xi^{-1}x_{\xi\tau}F_{n-2}\right)
+x_\xi^{-(p+n-1)}\p_\xi\left(x_\xi^{-1}F_{n-3,\xi}\right)
\end{align}
The equation is apparently an ODE in $\tau$. It may be easily
integrated, producing a first integral $C_n(\xi)$ at each order
$n$.

4. and 5. Apart of the above, for each pole the recurrence
relations have one resonant index $n=1-p$, where the l.h.s. turns
into zero. This resonance is compatible provided that the r.h.s.
is zero for this value of $n$. Thus the system of recurrence
relations \eqref{rec} is always over\-deter\-mined. If the number
of poles in the upper complex half-plane is $m$, then it involves
the coefficients ${F^j}_{-p-1}$ and the trajectories
$x^j,~j=1,...,m$, in $3m$ relations: $m$ dispersion relations
\eqref{dispL}, $m$ recurrence relations at $n=-p$, and $m$
compatibility conditions imposed on the recurrence relations at
$n=-p+1$ (the $p$ might be different for each pole, which we here
skipped for clarity of notation). After elimination of the
coefficients ${F^j}_{-p-1}$, this yields $2m$ constraints on $m$
trajectories $x^j(\xi,\tau)$, which in turn require $m$
compatibility conditions. When all these conditions are satisfied,
the other coefficients are uniquely determined by the recurrence
relations.

Once we get the singularity trajectories $x^j(\xi,\tau)$, we do
not have to sum up the Laurent series to obtain the solution. The
dispersion relation yields the acceleration
$a(x,t)=x_{\tau\tau}\left(\xi\left(x,t\right),\tau\left(x,t\right)\right)$.
When $a$ is known, the Vlasov equation becomes linear and we can
solve it readily by the method of characteristics. This makes the
extension of the P-test a potential source of new solutions to the
Vlasov equation.

The singularity analysis of the Vlasov equation provides
classification of its solutions by the number and multiplicity of
the poles. On the other hand, it does not apply to the solutions
with no poles, like the Maxwellian velocity distribution.

In uncomplicated cases the singularity analysis helps to find the
solutions explicitly. Below we consider the case of simple (first
order) poles.

\subsection{Simple poles}

When all poles are simple, the integrated recurrence relations
\eqref{rec} read for each pole (whose index $j$ is omitted)
\begin{align}\label{rec1}
(2-n){x_\xi}^{2-n}x_{\xi\tau}F_{n-1}=&-{x_\xi}^{2-n}F_{n-2,\xi}
-(3-n)\int_0^\tau
d\tau'\,{x_\xi}^{2-n}\p_\xi({x_\xi}^{-1}x_{\xi\tau}F_{n-2})~~~~\nn\\
&+\int_0^\tau
d\tau'\,{x_\xi}^{2-n}\p_\xi({x_\xi}^{-1}F_{n-3,\xi})+C_n(\xi)
\end{align}
where all the functions in the integrands are calculated at
$(\xi,\tau')$.

It follows from \eqref{rec1} that all the constraints are imposed
on ${F^j}_0$. This way, if we have $m$ poles, ${F^j}_0$ are given
by
\bi\item $m$ dispersion relations \eqref{dispL};
\item
$m$ recurrence relations \eqref{rec1} for $n=1$ (the superscript
$j$ at all the variables and the subscript 1 at $C$ are omitted)
\be\label{rec1a}
x_\xi x_{\xi\tau}F_0=C(\xi);
\ee
\item
$m$ compatibility conditions for the recurrence relation at the
resonant index $n=2$. Each condition may be integrated once over
$\xi$ to yield (also without the superscript~$j$)
\be\label{rec1b}
[\p_\tau+{x_\xi}^{-1}x_{\xi\tau}]F_0=Z(\tau),
\ee
where $Z$ is a $\tau$-dependent first integral.
\ei
The system (\ref{dispL},\ref{rec1a},\ref{rec1b}) consists of
$3m$ equations with $2m$ unknown functions: $m$ functions
${F^j}_0$ and m Lagrangian trajectories $x^j$. Hence it has to
satisfy $m$ mixed-derivative compatibility conditions.

Calculation of the compatibility conditions, which is
straightforward for linear equations, requires more effort for
nonlinear PDE. In this article we concentrate on the case of one
simple pole.

\subsection{Solution for one simple pole}
In this case the system (\ref{dispL},\ref{rec1a},\ref{rec1b})
consists of 3 equations with two unknown functions. After a
cumbersome, though straightforward calculation, we obtain a
condition of the form
\be\label{cond1}
\sqrt{C(\xi)}A(x_\tau,T(\tau))=0.
\ee
In \eqref{cond1}, $T$ is a $\tau$-only dependent first integral,
while $A$ is an algebraic expression in its variables. Hence
either $C=0$, or $x_\tau$ is a function of $\tau$ only,
independent of $\xi$. For $C=0$, equation \eqref{rec1a} also
requires $x_{\xi\tau}=0$, except for the trivial case of no
electric field, i.e. stationary solution $x_{\tau\tau}=0$. There
is one exception, due to the fact that $x_\tau$ occurs in
\eqref{cond1} accompanied by the factor $Z(\tau)$ or its
derivatives: if $Z\equiv 0$, then the variable $x_\tau$ vanishes
from the expression \eqref{cond1} and the condition is imposed on
the first integrals only. Further straightforward calculation
shows that in this case $x_{\tau\tau}$, although not necessarily
zero, has to be independent of $\xi$ (uniform electric field, 
which entails local neutrality of the plasma).
Eventually we obtain the general solution for $x(\xi,\tau)$ (and
consequently for the singularity trajectories $\ph=x_\tau$). If we
impose the initial condition $x(\xi,0)=\xi$, the solution reads
\be\label{sol-x}
x(\xi,\tau)=\xi u + 2\pi i \om_p^2 v_0\tau_0^3\left[1+
u\,(\ln\,u-1)\right]+v_1\tau,
\ee
where $u=1+\tau/\tau_0$ while $\tau_0$, $v_0$ and $v_1$ are
arbitrary constants. The constant $v_0$ arises as the value of the
first integral $C(\xi)$ defined in \eqref{rec1a}. Its reduction to
a constant is a necessary condition of compatibility. An example
of such trajectories is shown in Fig. 1.
\begin{figure}\label{traj}
\begin{center}
\includegraphics[width=0.9\textwidth]{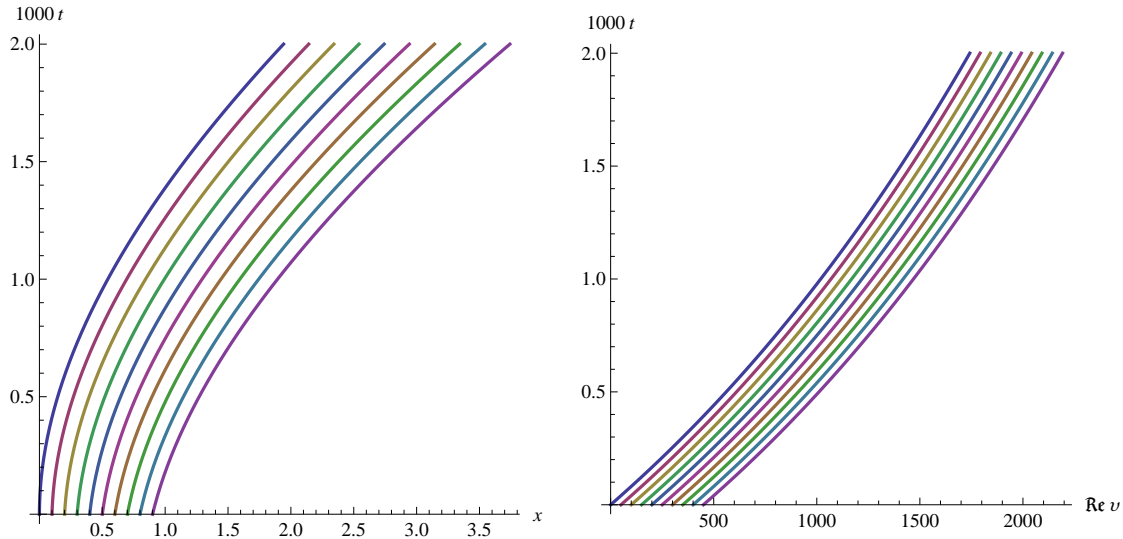}
\end{center}
\caption{{\bf Example of the trajectories:} The left diagram shows
the trajectories $x(\xi,\tau)$, the right one the corresponding
real parts of the singularities $v=\ph(\xi,\tau)$ for incrementing
values of the initial position $\mathop{\mathfrak{Re}}\xi$. The
other parameters are $v_0=-i,~ v_1= 0.001,~ \tau_0 =0.002,~ \om_p
=10^4$. The imaginary part of the velocity
$\mathop{\mathfrak{Im}}(\xi)/\tau_0$ remains constant throughout
the evolution. }
\end{figure}

The corresponding singularity manifold $v-\ph(\xi,\tau) =0$ may be
obtained from $\ph=x_\tau$
\be\label{ini-sing}
\ph(\xi,\tau)=\xi/\tau_0+2\pi i \om_p^2v_0\tau_0^2\ln u+v_1
\ee
The acceleration due to the electric field is fading
\be\label{accel}
a(x,\tau)= x_{\tau\tau}\bigl(\xi(x,t),\tau\bigr)=\frac{2\pi
i\om_p^2v_0\tau_0^2}{\tau_0+\tau}.
\ee
It is independent of $\xi$ (and consequently of $x$) which means
that the electric field is spatially uniform all the time.

Finally, the residuum $F_0$ of the solution $F$, at the pole
surface $v=\ph(\xi,\tau)=\phi(x,t)$, is equal to $a(x,t)$, up to a
constant factor
\be
F_0(\xi,\tau)=v_0 \tau_0^2/(\tau_0 + \tau)
\ee
The singularity at $\tau=-\tau_0$ does not belong to the family of
those solvable with respect to $v$. It is harmless if $\tau_0>0$
and we consider only solutions for
$\mathop{\mathfrak{Re}}(\tau)>0$. The calculation is local in
$\xi$ and $\tau$ and may leave $\tau=\tau_0$ outside the
considered singularity manifold.

The solution for $F$ may be obtained on the basis of the
$x_{\tau\tau}$ calculated in \eqref{accel}, as the acceleration is
the same $a(x,t)$ for the singularity trajectory and all the other
electron trajectories. By the method of characteristics we obtain
in a straightforward way
\be\label{sol-F}
F(x,v,t)=F\bigl(x\!-\! v t\!-\!2\pi i
\om_p^2\tau_0^3v_0\,[\ln(1\!+\!t/\tau_0)-t/\tau_0],\,v\!-\!2\pi i
\om_p^2\tau_0^2 v_0\ln(1\!+\!t/\tau_0) ,0\bigr),
\ee
assuming that the initial conditions are set out at $t=\tau=0$
(please note that the constants $x_0$, $\tau_0$ and $v_0$ are not
the initial values of $x$, $\tau$ and $x_\tau$).

To comply with the assumption of the first order pole, we have to
limit the initial conditions to those having a pole at
$v=\ph(x,0)$ in the upper half-plane. The physical sense of
$F(x,v,t)$ requires a symmetric (complex conjugate) second pole in
the lower half-plane. An example of such initial conditions could
be the Cauchy distribution in $v$
\be
\frac {1}{\pi} \,\frac{\gamma}{(v-v_C)^2+\gamma^2},
\ee
where the center-of-mass velocity $v_C=0$ if the plasma as a whole
is at rest, while $\pm i\gamma$ are positions of the poles. Then
$\gamma=\pm\mathop{\mathfrak{Im}}(\xi/\tau_0)$.

The initial spatial dependence of $F$ is not limited in principle, except 
that the integral of $F$ over velocity has to be independent of $x$, 
to ensure uniformity of the electric field. The
physical sense of $F$ as the cumulative distribution in $x$
requires that it be a positive and non-decreasing function of $x$ 
for all $v$. It follows from \eqref{sol-F} that an initially uniform 
(in space) distribution $F$ remains
spatially uniform throughout the evolution. Thus no spontaneous
breaking of the translational symmetry occurs.

\section*{Conclusions}
The singularity analysis may be performed for the Vlasov equation,
though the analysis is not the typical P-test.

For the one-dimensional Vlasov equation, the method provides
classification of solutions which are globally meromorphic in the
velocity space and free of movable branching in the position and
time, except for the constant-time singularity $t=-\tau_0$.

The common acceleration $a(x,t)$ at a given point $x$ and moment
$t$ can be calculated, provided that we find the trajectories of
the poles $v={x^j}_{\tau^j}(\xi^j,\tau^j)$, where $\xi^j,~\tau^j$
are the Lagrangian coordinates given by \eqref{Lagr}. Then the
solution of the Vlasov equation may be obtained by the method of
characteristics.

The system of equations determining the Eulerian from Lagrangian
coordinates is an overdetermined system. The solutions satisfying
the P-test are selected by the compatibility condition.

In case of simple poles, all the constraints are imposed on the
lowest-order coefficient ${F^j}_0(\xi^j,\tau^j)$.

The simplest case of one simple pole in the upper complex
half-plane ($\mathop{\mathfrak{Im}}{v}>0$) may easily be solved
explicitly. The solution corresponds to a uniform electric field.
An initially uniform solution remains uniform throughout the
evolution, however the method works also for nonuniform initial
distributions. In all these cases the characteristics are free
from crossings, thus the initial position of the trajectory can be
uniquely determined, given their actual position and time, from
equation \eqref{Lagr}.

Obviously the case of the uniform electric field could have been
solved without the P-test machine. However the method also works
for more complex situations, with greater number of poles. The
compatibility condition provides an extra equation, which
restricts the class of solutions and may facilitate solving the
equations. The case of more than one pole will be discussed in the
next paper.

The method requires that the solution has poles in the velocity
space. Therefore it does not cover the important Maxwellian
distribution.

\end{document}